\documentstyle[aps]{revtex}
\def \be   {\begin{equation}}
\def \ee   {\end{equation}}
\def \l {\label}
\begin{document}
\input epsf
\baselineskip=25pt
\title{Discrete scalar field and general relativity}
\author{Manoelito M de Souza\footnote{E-mail: manoelit@cce.ufes.br}}
\address{Departamento de
F\'{\i}sica - Universidade Federal do Esp\'{\i}rito Santo\\29.065-900 -Vit\'oria-ES-Brazil}
\date{\today}
\maketitle
\begin{abstract}
\noindent What is the nature - continuous or discrete - of matter and of its fundamental interactions? The physical meaning, the properties  and the consequences of a discrete scalar field are discussed; limits for the validity of a mathematical description of fundamental physics in terms of continuum fields are a natural outcome of discrete fields with discrete interactions. Two demarcating points (a near and a far) define a domain where  no difference between the discrete and the standard continuum field formalisms can be experimentally detected. Discrepancies, however, can be observed as a  continuous-interaction is always stronger below the near point and weaker above the far point than a discrete one. The connections between the discrete scalar field and gravity from general relativity are discussed. Whereas vacuum solutions of general relativity can be retrieved from discrete scalar field solutions, this cannot be extended to solutions in presence of massive sources as they require a true tensor metric field. Contact is made, on passing, with the problem of dark matter and the rotation curve of galaxies, with inflation and the accelerated expansion,  the apparent anomaly in the Pioneer spacecraft acceleration,  the quantum Hall effect, high-$T_{c}$ superconductivity, quark confinement, and with Tsallis generalized one-parameter statistics as some possible manifestation of discrete interaction and of an essentially discrete world.
\end{abstract}
\begin{center}
\end{center}
\section{Introduction}
Although a scalar field has not been observed in nature as a fundamental field its use as such is very frequent in the modern literature, particularly in elementary particles, field theory and cosmology. Here we will apply to the scalar field the concepts and results developed in the reference \cite{hep-th/0006237}, where the concept of a discrete field was introduced and its wave equation and its Green's function discussed. The standard field and its formalism, which for a distinction, we always append the qualification continuum, are retrieved from an integration over the discrete-field parameters. Remarkable in the discrete field is that it has none of the problems that plague the continuum one so that the meaning and origin of these problems can be left exposed on the passage from the discrete to the continuum formalism. Although the motivations for the introduction of a generic discrete field in \cite{hep-th/0006237} have being made on pure physical grounds of causality, a deeper discussion about its physical interpretation have been left for subsequent papers on specific fields. This discussion will be retaken here with the simplest structure of a field, the scalar one. The idea of a pointlike field, although unusual, represents the same symmetry of quantum field theory where fields and sources are equally treated as quantized fields. Here it is seen from a classical perspective. Besides, pointlike object is not a novelty in physics and one of the major motivations of the, nowadays so popular, string theory is of avoiding \cite{Polchinski} infinities and acausalities in the fields produced by point sources, problems that do not exist for the discrete field \cite{hep-th/9610028}. 

This paper is structured in the following way. Section II, on the sake of a brief review of the mathematical definition of discrete fields, is a recipe on how to pass from a continuum to a discrete field formalism, and vice-versa. The discrete scalar field, its wave equation, Lagrangian and energy tensor are briefly discussed in Section III. Section IV discusses the consequences of discrete interactions for the mathematical description of the physical world. Then it gains generality as the discussions leaves the specificity of scalar interactions widening to the universality of all fundamental interactions. Calculus (integration and differentiation) which is based on the opposite idea of smoothness and continuity, has its full validity for describing dynamics restricted then to a very efficient approximation  in the case of a high density of interaction points, such that the concept of acceleration as a continuous change of velocity may be introduced in an effective physical description of fundamental interactions. This seems to be an answer to the Wigner's pondering \cite{Wigner} about the reasons behind the unexpected effectiveness of mathematics on the physical description of the world.  In Section V the discrete scalar field reveals and unavoidable connection to gravity of general relativity as both have energy-momentum for sources. Besides, it has been shown \cite{gr-qc/9801040} that the gravitational field of general relativity in a vacuum is reduced to an effective spacetime manifestation of a discrete scalar field. We will see here that this cannot be extended to gravity in the presence of massive sources; a true second-order tensor field is required.
 The reader should be aware, however, that this is not a paper about a new theory of gravity; it is about the meaning and implications of an assumed discreteness of all matter and interactions, particularly of the scalar ones. Its connection to gravity is a consequence. Theoretical implications of discrete interactions and the possibility that some of its consequences may have already been observed are considered in the concluding Section VI. 

\section{From continuum to discrete}

For a concise introduction of the discrete-field concept it is convenient to replace the Minkowski spacetime flat geometry by a conical projective one in an embedding (3+2) flat spacetime: 
\be
\l{def}
\{x\in R^4\}\Rightarrow\{x,x^5\in R^5{\big|}(x^5)^2+x^2=0\},
\ee
where $x\equiv({\vec x},t)$ and $x^2\equiv\eta_{\mu\nu}x^{\mu}x^{\nu}=|{\vec x}|^2-t^2.$ So a change $\Delta x^5$ on the fifth coordinate, allowed by the constraint $(\Delta x^5)^2+(\Delta x)^2=0,$   is a Lorentz scalar that can be interpreted as a change $\Delta\tau$ on the proper-time  of a physical object propagating across an interval $\Delta x:\quad \Delta x^5=\Delta\tau=\pm\sqrt{(\Delta t)^2-(\Delta{\vec x})^2}.$ 
The constraint
\be
\l{hcone}
(\tau-\tau_{0})^2+(x-x_{0})^2=0
\ee
defines a double hypercone with vertex at $(x_{0},\tau_{0}),$ whilst
\be
\l{hplane}
(\tau-\tau_{0})+f_{\mu}(x-x_{0})^{\mu}=0
\ee
defines a family of hyperplanes tangent to the double hypercone and labelled by their normal\footnote{The Eq. (\ref{hplane}) can be written in $R^5$ as $f_{M}\Delta x^{M}=0,\;\;M=1,2,3,4,5$ with $f_{M}=(f_{\mu},1)$} $f_{\mu}$, a constant four-vector. The intersection of the double hypercone with a hyperplane defines its $f$-generator tangent to $f^{\mu}$  ($f^{\mu}:=\eta^{\mu\nu}f_{\nu}).$ The projective conical geometry (\ref{def}) allows that a propagating continuum field be defined in a most natural way as the hypercone (\ref{hcone}) is its very manifold support
\be
\l{phi}
\phi(x)\Rightarrow\phi(x,\tau){\Big|}_{\Delta\tau^2+\Delta x^2=0},
\ee
which is just the expression of local causality as contained in the standard Lienard-Wiechert solutions \cite{Jackson}, for a known example with $\Delta\tau=0$ (massless field). But the projective geometry (\ref{def}) allows a further step of introducing the new concept of discrete  field just by restricting, in a manifestly covariant way, the field support to the  straight line intersection of the hypercone (\ref{hcone}) with its tangent plane (\ref{hplane}) (extended causality).
\be
\l{df}
\phi_{f}(x,\tau):=\phi(x,\tau){\Big|}_{{\Delta\tau+f.\Delta x=0}\atop{\Delta\tau^2+\Delta x^2=0}}:=\phi{\Big|}_{f}.
\ee
This makes the great difference. The discrete field has discrete, pointlike wave fronts, in contraposition to the (characteristics) 2-surface  wave fronts of the continuum field. 
The symbol ${\big|}_{f}$ is a short notation for the double constraint in the middle term of Eq. (\ref{df}). The constraint (\ref{df}) induces the directional derivative (along the fibre $f$, the hypercone generator $f$)
\be
\l{dd}
\nabla_{\mu}\phi_{f}(x,\tau):=(\partial_{\mu}-f_{\mu}\partial_{\tau})\phi_{f}(x,\tau).
\ee
An action for a discrete scalar field is
\be
\l{da1}
S_{f}=\int_{hc} d^5x{\Big\{}\frac{1}{2}\eta^{\mu\nu}\nabla_{\mu}\phi_{f}(x,\tau)\nabla_{\nu}\phi_{f}(x,\tau)-\chi\phi_{f}(x,\tau)\rho(x,\tau){\Big\}},
\ee
where $d^5x=d^4xd\tau$, $\chi$ is a coupling constant, $\rho(x,\tau)$ is the source for the scalar field, and $hc$ stands for hypercone, the integration domain. It can also be written  as
\be
\l{ar}
S_{P}=\int_{hc} d^{4}xd\tau\delta(\Delta\tau+f.\Delta x){\Big\{}\frac{1}{2}\eta^{\mu\nu}\partial_{\mu}\phi\partial_{\nu}\phi-\chi\phi(x,\tau)\rho(x,\tau){\Big\}},
\ee
as the very  restriction to the hyperplane (\ref{hplane}) by itself implies on (\ref{da1}). $P$ stands for any generic fixed point, a local hypercone vertex: $P=(x_{0},\tau_{0}),$ $\Delta\tau=\tau-\tau_{0}$ and $\Delta x=x-x_{0}.$ Local causality, dynamically implemented through the field equation solutions, implies that the field propagates on a hypercone (lightcone, if a massless field) with vertex on $P,$  which is an event on the world  line of $\rho(x,\tau)$. The constraint (\ref{hplane}), included in this action (\ref{ar}), further restricts the field to the fiber $f$, expressing an extended concept of causality \cite{hep-th/0006214,hep-th/0006237}.
There is no mass term in this Lagrangian and, nonetheless, as discussed  in \cite{hep-th/0006237}, it still describes both, massive and massless fields. The mass of a massive discrete field is implicit on its propagation with a non-constant proper-time, $\Delta\tau\not=0$. Eq. (\ref{da1}) is a scale-free action expressing the (1+1)-dynamics of a discrete field, massive or not, on a fibre $f$; a mass term would break its conformal symmetry \cite{hep-th/0006237}. 

Then the field equation and the tensor energy for a discrete field are, respectively,
\be
\l{dfe}
\eta^{\mu\nu}\nabla_{\mu}\nabla_{\nu}\phi_{f}(x,\tau)=\chi\rho(x,\tau),
\ee
\be
\l{det}
T^{\mu\nu}_{f}=\nabla^{\mu}\phi_{f}\nabla^{\nu}\phi_{f}-\frac{1}{2}\eta^{\mu\nu}\nabla^{\alpha}\phi_{f}\nabla_{\alpha}\phi_{f}.
\ee
Eqs. (\ref{da1},\ref{dfe},\ref{det}) must be compared to the standard expressions for the continuum field; their meaning and validity will be discussed in Section IV. The passage from a continuum to a discrete field formalism can be summarized in the following schematic recipe (the arrows indicate replacements):
\be
\l{recipe}
\cases{\{x\}\Rightarrow\{x,x^5\};\cr
\phi(x)\Longrightarrow \phi(x,\tau){\Big |}_{f};\cr
\partial_{\mu}\Rightarrow\nabla_{\mu},\cr}
\ee
accompanied by a dropping of the mass term from the Lagrangian. Moreover a discrete field requires, for consistency, a discrete source. A continuous $\rho(x)$ is replaced by a discrete set of pointlike sources $\rho(x,\tau)$. Any apparent continuity is reduced to a question of scale in the observation. $\rho(x,\tau)$ is, like  $\phi_{f}(x,\tau)$, a discrete field defined on a hypercone generator too, which just for simplicity, is  not being considered here. This is a symmetry between fields and sources: they are all discrete fields, and the current density of one is the source of the other.

Reversely, in the passage from discrete to continuum, the continuum field and its field equations are recuperated in terms of effective average fields smeared over the hypercone
\be
\label{s}
\phi(x,\tau)=\frac{1}{2\pi}\int_{hc} d^{4}f\phi_{f}(x,\tau).
\ee
A cone is the union of all its generators; a continuum field, which has support on a hypercone, is retrieved from a discrete field by integrating out its $f$-parameter which labels the hypercone-generators.
The discrete-to-continuum passage in Eq. (\ref{s}) provokes the appearing of the mass term and the breaking of the conformal symmetry of the action (\ref{da1}). More specifically, integration over the $f$-parameters reproduces from Eq. (\ref{dfe}) and from Eq. (\ref{pr9}) below, respectively the standard  wave equation and the Green's function of the continuous formalism, including the mass term, in the case of a massive field. This has been explicitly proved, for both the massive and the massless fields, in the references \cite{hep-th/0006237,gr-qc/9801040}. Here we discuss their physical implications. 

For intense continuum fields there is, a not exact but insightful, picture of the discrete field $\phi_{f}(x)$ as the part of the wave front of $\phi(x)$ that is propagating along a generator $f$ of its hypercone. Then, it would be the restriction of $\phi(x)$ to $f$:
\be
\phi_{f}(x,\tau)=\phi(x){\Big|}_{f}.
\ee
It makes more visible how an integration over the parameter $f$ of a discrete field retrieves the continuum one. But this is, anyway, just an approximated heuristic picture whose limitation can be exposed by considering the opposite extreme case of a very weak field constituted by just a single quantum (discrete field) exchanged. Then it is clear that Eq. (\ref{s}) represents rather the smearing of the single discrete field onto  a hypercone section. Replacing $\phi_{f}$ by $\phi$ is to describe the physical content (momentum, energy, charge, etc) that exists at a single space point at a time as if it were distributed over a hypercone (spherical, if isotropic) 2-section. This has been discussed in \cite{gr-qc/9801040}

\section{The discrete scalar field}
Whereas there is no restriction on $\rho(x)$ for a continuum field, for a discrete one, as already mentioned, it must be a discrete set of point sources. A continuously extended source would not be consistent as it would produce a continuum field. The source of a discrete scalar field is then given by
\be    
\l{rof}
\rho(x,\tau_{x})=q(\tau_{z})\delta^{(3)}({\vec x}-{\vec z}(\tau_{z}))\delta(\tau_{x}-\tau_{z}),
\ee
where $z(\tau)$ is its world  line parameterized by its proper time $\tau$; $q(\tau)$ is the scalar charge whose physical meaning will be made clear later. The sub-indices in $t$ and $\tau$ specify the respective events $x,\;y$ and $z$. 
The field Eq. (\ref{dfe}) is solved by\footnote{There is a slight change of notation with respect to \cite{hep-th/0006237}: $G_{f}\Rightarrow\frac{dG_{f}}{d\tau}$.}
\be
\l{phif}
\phi_{f}(x,\tau)=\int d^5y\frac{d}{d\tau_{y}}G_{f}(x-y,\tau_{x}-\tau_{y})\rho(y,\tau_{y})
\ee
with
\be
\l{boxGf}
\eta^{\mu\nu}\nabla_{\mu}\nabla_{\nu}\frac{d}{d\tau}G_{f}(x,\tau)=\delta^{(5)}(x)=\delta(\tau)\delta^{(4)}(x).
\ee
The discrete Green's function associated to the Klein-Gordon operator is given\cite{hep-th/0006237} by
\be
\label{pr9}
\frac{d}{d\tau}G_{f}(x,\tau)=\frac{1}{2}\theta(bf^4t)\theta(b\tau)\delta(\tau+ f.x),\;\;\;\;{\vec x}_{{\hbox{\tiny T}}}=0,
\ee
where $b =\pm1,$ $f^4$ is the fourth component of $f^{\mu}$, and $\theta (x)$ is the Heaviside function, $\theta(x\ge0)=1$ and $\theta(x<0)=0.$ The labels {\tiny L} and {\tiny T} are used as an indication of, respectively,  longitudinal and transversal with respect to the space part of $f$: ${\vec f}.{\vec x}_{\hbox{\tiny T}}=0$ and $ x_{{\hbox{\tiny L}}}=\frac{{\vec f}.{\vec x}}{|{\vec f}|}$. We observe that $G_{f}(x,\tau)$ is finite, like a product of three $\theta$-functions. 
Remarkably it does not depend on anything outside its support, the fibre $f$, as stressed by the append ${\vec x}_{{\hbox{\tiny T}}}=0$. One could retroactively use this knowledge in the action (\ref{da1}) for rewriting it as
\be
\l{sf3}
S_{f}=\int_{hc} d^{5}x\delta^{(2)}({\vec x}_{\hbox{\tiny T}}){\Big\{}{1\over2}\eta^{\mu\nu}\nabla_{\mu}\phi_{f}\nabla_{\nu}\phi_{f}-\chi\phi_{f}(x,\tau)\rho(x,\tau){\Big\}},
\ee
just for underlining that the fibre $f$ induces a conformally invariant (1+1) theory of massive and massless fields, embedded in a (3+1) theory, as generically discussed in \cite{hep-th/0006237}. Actually, the factor $\delta^{(2)}({\vec x}_{\hbox{\tiny T}})$ is an output of the actions (\ref{da1}) or (\ref{ar}) (it is not necessary to put it in there by hand) and it can never be incorporated as a factor in the definition (\ref{pr9}) of $G_{f}(x,\tau)$, except under an integration sign as in Eqs. (\ref{phif}) and (\ref{sf3}). Then one could, just formally, use
\be
\l{dsf}
\rho_{[f]}(x-z,\tau_{x}-\tau_{z})=q(\tau)\delta(\tau_{x}-\tau_{z})\delta(t_{x}-t_{z})\delta(x_{\hbox{\tiny L}}-z_{\hbox{\tiny L}}),
\ee
where $\rho_{[f]}$ represents the source density $\rho$ stripped of its explicit ${\vec x}_{\hbox {\tiny T}}$-dependence, for reducing the action to  
\be
\l{21}
S_{f}=\int d\tau_{x}dt_{x}dx_{\hbox{\tiny L}}{\Big\{}{1\over2}\eta^{\mu\nu}\nabla_{\mu}\phi_{f}\nabla_{\nu}\phi_{f}-\chi\phi_{f}(x,\tau)\rho_{[f]}(x,\tau){\Big\}},
\ee
by just omitting the irrelevant transversal coordinates. Eq. (\ref{da1}) then, after its output Eq. (\ref{pr9}), is formally equivalent to Eq. (\ref{21}). But we should observe that this is no more than a formal expression once $\phi_{f}\rho_{[f]}$ is not null at just one event, the intersection of the worldline of $\rho(x)$, whose support is not $f$, with the fibre $f$, support of $\phi_{f}(x)$. See the Figure 1.

\vglue13cm
\hglue-1.0cm
\begin{minipage}[]{7.0cm}
\parbox[t]{5.0cm}{
\begin{figure}
\vglue-13cm
\epsfxsize=200pt
\epsfbox{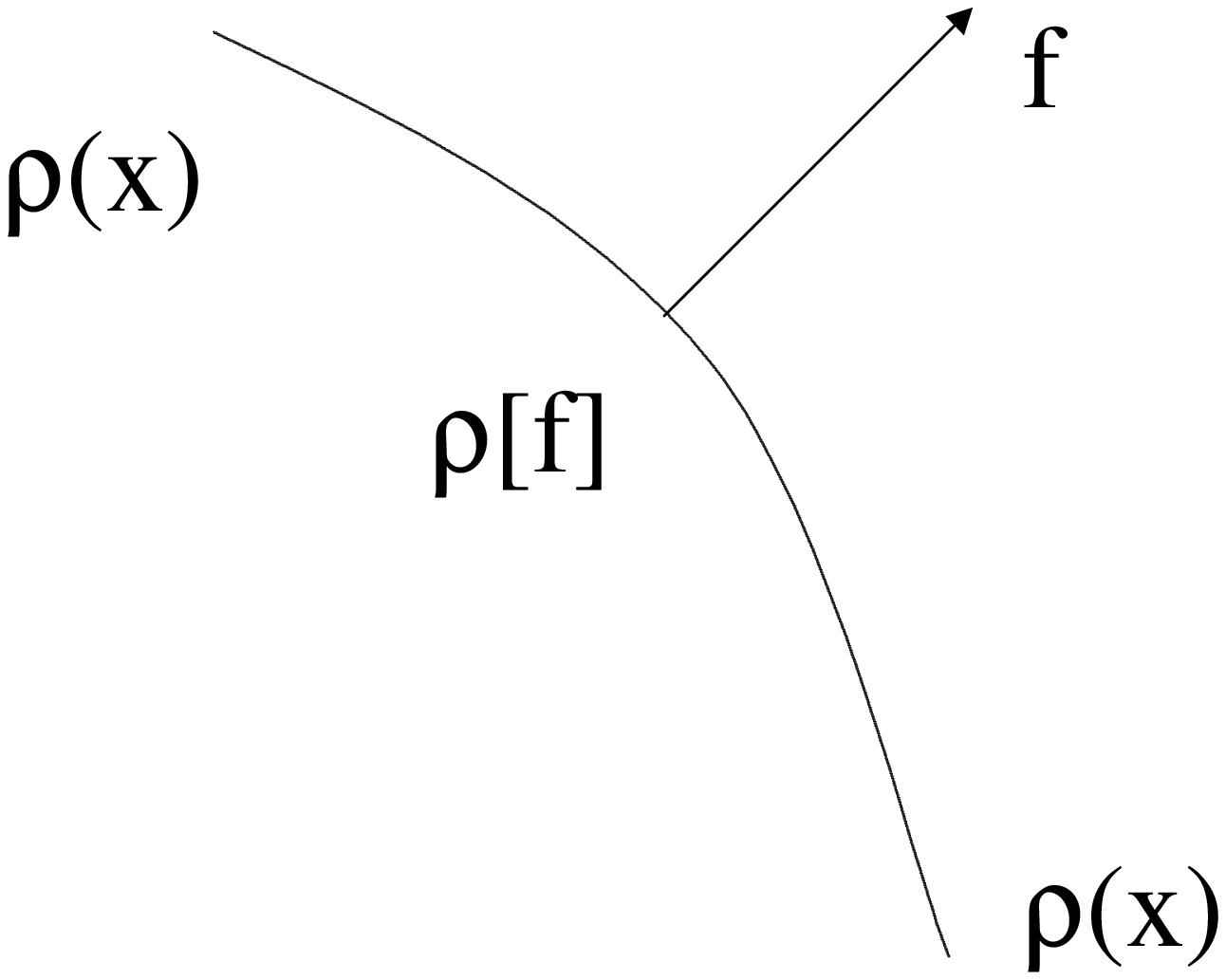}
\end{figure}}
\end{minipage}\hfill
\vglue-7cm
\hglue-1.0cm
\begin{minipage}[]{7.0cm}
\begin{figure}
\hglue7.0cm
\parbox[t]{5.0cm}{\vglue-6cm
\caption{The meaning of $\rho_{[f]}$: the value of $\rho(x)$ at the specific point defined by the intersection of the worldline of $\rho(x)$ with the fibre $f$, support of $\phi_{f}(x)$. }}
\end{figure}
\end{minipage}

\vglue-1cm

The solutions from the standard wave equation with $m=0$, for a point source, are well known spherical waves propagating (forwards or backwards in time) on a lightcone in contradistinction to the solutions (\ref{pr9}) that are projections on a lightcone of, massive or massless point signals propagating {\it always forwards in time} on a straight line, a generator of the hypercone (\ref{hcone}). 
With $b=+1$ and $f^4\ge1$ which implies an  emitted field, one has from Eqs. (\ref{pr9}) and (\ref{rof}) that
$$
\phi_{f}(x,\tau_{x})=\chi\int d^{5}y \theta(t_{x}-t_{y})\theta(\tau_{x}-\tau_{y})\delta[\tau_{x}-\tau_{y}+ f.(x-y)]q(\tau_{z})\delta^{4}(y-z)=$$
\be
=\chi\int d\tau_{y} \theta(t_{x}-t_{z})\theta(\tau_{x}-\tau_{y})\delta[\tau_{x}-\tau_{y}+ f.(x-z)]q(\tau_{z}),
\ee
where an extra factor 2 accounts for a change of normalization with respect to Eq. (\ref{pr9}) due to the exclusion of the annihilated field (which corresponds in Eq. (\ref{s}) to the integration over the future lightcone). Being massive or massless is determined by $\tau$ being constant or not. For a massive field, its mass and its timelike four velocity are hidden behind its lightlike projection $f$ and its non-constant $\tau$; they  become explicit only after the passage from discrete to continuum fields. But as it will be made clear in Section V, there is no point on considering a massive discrete scalar field because any discrete scalar field must be associated to the gravitational field of general relativity. So massive discrete scalar fields will not be considered here any further; $\Delta\tau\equiv0,\;\theta(\Delta\tau)\equiv1$.
Then,
\be
\l{Af0}
\phi_{f}(x_{{\hbox{\tiny L}}},{\vec x}_{{\hbox{\tiny T}}}={\vec z}_{{\hbox{\tiny T}}},t_{x},\tau_{x}=\tau_{z})=\chi \theta(t_{x}-t_{z})q(\tau_{z}){\Big |}_{f.(x-z)=0}
\ee
or for short, just
\be
\l{Af1}
\phi_{f}(x,\tau)=\chi q(\tau)\theta(t){\Big |}_{f}.
\ee
$\nabla\theta(t)$ does not contribute to $\nabla \phi_{f},$ except at $x=z(\tau),$ as a further consequence of the field constraints. So, for $t>0$ one can write just 
\be
\l{AfV}
\phi_{f}(x,\tau_{x})=q(\tau_{z}){\Big |}_{f}
\ee
\be
\l{dAf}
\nabla_{\nu}\phi_{f}=-f_{\nu}\frac{dq}{d\tau_{z}}{\Big |}_{f}=-f_{\nu}{\dot q}{\Big |}_{f}
\ee 
With Eq. (\ref{dAf}) in Eq. (\ref{det}) one has
\be
\l{tsf1}
T^{\mu\nu}_{f}(x,\tau_{x})=f^{\mu}f^{\nu}{\dot q}^{2}{\Big |}_{f}
\ee
The field four-momentum, given by $\int T^{\mu\nu}n_{\nu}d\sigma$ for a continuum field, is reduced, thanks to the field pointlike character and to its independence from the transversal coordinates, to
\be
\l{psf}
p^{\mu}_{f}=T^{\mu\nu}_{f}n_{\nu}=f^{\mu}{\dot q}^{2}{\Big |}_{f}
\ee
where $n$ is a spacelike four vector such that $n.f=1$. The conservation of the energy-momentum content of $\phi_{f}$ is assured then just by $f$ being lightlike, $f^2=0,$
\be
\l{emc}
\nabla_{\mu}T^{\mu\nu}_{f}=-2f_{\mu}f^{\mu}f^{\nu}{\dot q}{\ddot q}{\Big |}_{f}=0.
\ee
It is justified naming $\phi_{f}$ a discrete field because although being a field it is not null at just one space point at a time; and it is not a distribution, a Dirac delta function, as it is everywhere and always finite. Its differentiability, in the sense of having space and time derivatives, is however assured by its dependence on $\tau$, a known continuous spacetime function. It is indeed a new concept of field, a very peculiar one, discrete and differentiable; it is just a finite pointlike spacetime deformation projected on a null direction $f$, with a well defined and everywhere conserved energy-momentum. It is this discreteness in a field that allows the union of wave-like and particle-like properties in a same physical object (wave-particle duality); besides this implies \cite{hep-th/9911233} finiteness and no spurious degree of freedom (uniqueness of solutions).

 \section{Discrete physics}

According to Eq. (\ref{AfV}), the field $\phi_{f}$ is given, essentially, by the charge at its retarded time, i.e. the amount of scalar charge at $z_{ret}$, the event of its creation.  It has a physical meaning, in the sense of having an energy-momentum content, if and only if ${\dot q}{\Big |}_{f}\not=0$. So, the emission or the absorption of a scalar field is, respectively, consequence or cause of a change in the amount of scalar charge on its source. This is so because emitting or absorbing a scalar field requires a change in the state of its source which is so poor of structure that has nothing else to change but itself, and this is fundamental for determining the scalar charge nature. The picture becomes clearer after recalling that we are dealing with discrete field and discrete interactions  which implies that the change in the state of a field source occurs at isolated events.  $q(\tau)$ is not a continuous function:  
\be
\l{dV}
q(\tau):=\sum_{i}q_{\tau_{i+1}}{\bar\theta}(\tau_{i+1}-\tau){\bar\theta}(\tau-\tau_{i}),
\ee
where
\be
\l{dth}
{\bar\theta}(x)=\cases{1, &if $x>0$;\cr 1/2,& if $x=0$;\cr 0,& if $x<0$,\cr}
\ee
and the index $i$ labels the interaction points on the source worldline, $i=1,2,3\dots$. For a given $\tau$ only one, or at most, two terms contribute to the sum in Eq. (\ref{dV})
\be
\l{qt}
q(\tau)=\cases{q_{\tau_{j}}, & if $\tau_{j}<\tau<\tau_{j+1}$;\cr
		&\cr
		\frac{q_{\tau_{j-1}}+q_{\tau_{j}}}{2}, & if $\tau=\tau_{j}$;\cr
		&\cr
		q_{\tau_{j-1}}, & if $\tau_{j-1}<\tau<\tau_{j}$,\cr}
\ee
as indicated in  the graph of the Figure 2.
\vglue13cm
\hglue-2.0cm
\begin{minipage}[]{7.0cm}
\parbox[t]{5.0cm}{
\begin{figure}
\vglue-14cm
\epsfxsize=300pt
\epsfbox{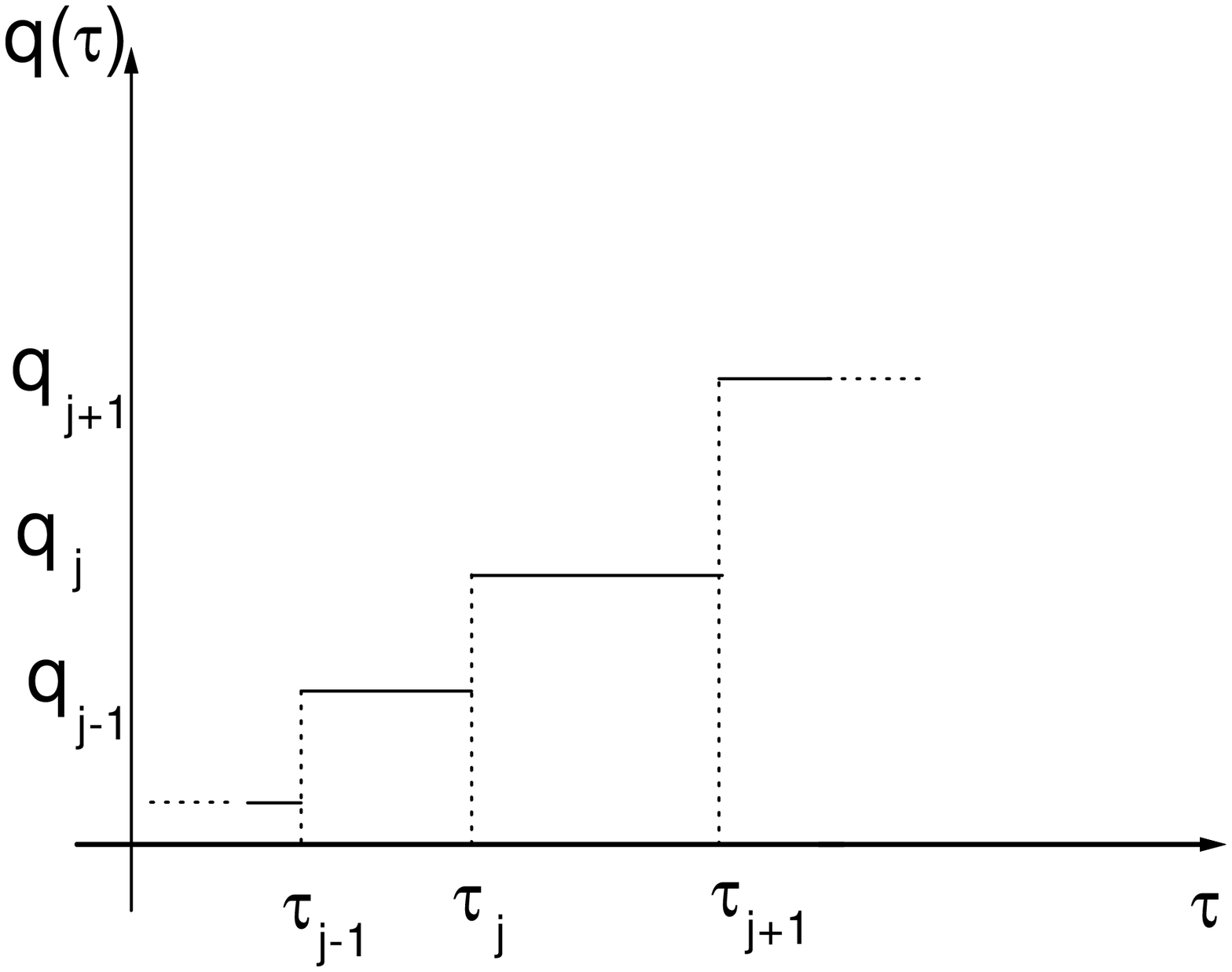}
\end{figure}}
\end{minipage}\hfill
\vglue-7cm
\hglue-1.0cm
\begin{minipage}[]{7.0cm}
\begin{figure}
\hglue9.0cm
\l{f3a}
\parbox[t]{5.0cm}{\vglue-6cm
\caption{Discrete changes on a discrete scalar charge along its worldline. A discrete scalar charge is so poor of structure that there is nothing else to change but itself. There is change in the state of a scalar source only at the interaction points on its worldline which is labelled by its proper time. In the limit of a worldline densely packed of interaction points a continuous graph is a good approximation.}}
\end{figure}
\end{minipage}

\vglue-1cm
The discretization introduced through Eqs (\ref{dV}) and (\ref{qt}) should not be mistaken with the discretization used to study field theory on a computer where the continuum spacetime is replaced by a grid and derivatives are replaced by ratios of finite differences. There it is just an approximative artifact. Here the spacetime is always a continuum and the discreteness is physically true; it stands for the absolute quantization of matter and all of its interactions.

The change in the state of the scalar source is not null only at the (discrete) interaction points and so, rigourously, it cannot  be defined as a time derivative, as there is no continuous variation, just a sudden finite change. The naive use of 
\be 
\l{dasdd}
{\dot q}=q(\tau)\delta(\tau-\tau_{z}),
\ee
would be just an insistence on an unappropriate continuum formalism, besides artificially introducing infinities where there is none. It means that one must replace time derivatives by finite differences 
\be
\l{finitedif}
{\dot q}(\tau)\Rightarrow \cases{\Delta q_{\tau_{j}} & if $\tau=\tau_{j}$;\cr 
				0 & if $\tau\neq\tau_{j}$,\cr}		
\ee
and a proper-time integration by a sum over the interaction points on the charge worldline. 
The existence and meaning of any physical property that corresponds to a time derivative must be reconsidered at this fundamental level. Velocity ($v)$ exists as a piecewise smoothly continuous function (discontinuous at the interaction points). Acceleration ($a$) and derivative concepts like force ($F$), power ($P$), etc rigorously do not exist. We must deal with finite differences, respectively, the  sudden changes of velocity ($v$), momentum ($p$) and energy ($E$):
\be
\l{reciped}
\cases{a\Rightarrow\Delta v\cr
F\Longrightarrow\Delta p \cr
P\Rightarrow\Delta E\cr}
\ee
For a large number of elementary (pointlike) interacting bodies $\Delta \tau_{j}$ is a statistical average time-interval between two consecutive interaction events on one body worldline. It decreases with the number of participants, and therefore, in the case of gravitational interaction, with the masses of the macroscopic interacting bodies.
The observability of an interaction discreteness is in fact controlled by the ratio of the two parameters shown in the Figure 2, $\Delta q_{j}$ and $\Delta\tau_{j},$ as the validity of an approximative continuum description of fundamental interactions requires the existence of the ratio 
\be
\l{ratio}
{\dot q}_{j}=
\frac{\Delta q_{j}\rightarrow0}{\Delta \tau_{j}\rightarrow0}\neq0,
\ee
which can then be interpreted as a time derivative of $q(\tau),$ taken as a smooth continuous function of $\tau$. But actually, in Eq. (\ref{ratio}) 
\be
\l{dX}
\Delta q_{j}\rightarrow0\qquad{\hbox{and}}\qquad\Delta \tau_{j}\rightarrow0
\ee
have the pragmatic meaning that both discrete changes, $\Delta q_{j}$ and $\Delta \tau_{j}$, are smaller than their respective experimental thresholds of detectability, which, of course, are existing-technology dependent. Therefore, $\frac{\Delta q_{j}}{\Delta\tau_{j}}$ would diverge if  $\Delta \tau_{j},$ but not $\Delta q_{j}$, would satisfy Eq. (\ref{dX}), and it would unduly\footnote{Because the actual interaction is not null.} be null in the case of only  $\Delta q_{j},$ but not $\Delta \tau_{j},$ satisfying it. The interaction intensity is described by ${\dot q}_{j}$ (as a time derivative of $q(\tau)$) for the continuous interaction but by the two parameters $\Delta q_{j}$ and $\Delta \tau_{j}$ for the discrete one.
 Both results, infinity and an undue zero, evince the existence of two demarcating point in the range of the ratio parameter $\frac{\Delta q_{j}}{\Delta \tau_{j}}$, warning for the inadequacy of the approximative continuous-interaction description: a near and a far point. These two  points delimit the range of the ratio-parameter $\frac{\Delta q_{j}}{\Delta\tau_{j}}$ where there is no observationally detectable difference between a discrete and a continuous interaction, defining the domain of validity of a continuum field as an effective physical description.  A continuum field is then stronger below the near point and weaker above the far one than its corresponding discrete field. Outside the range delimited by these points a discrete-interaction description must be used. This is schematically represented in Figure 3 with the variations of a probe scalar charge $q$ versus its distance $R$ from the origin, site of a central charge. The two graphs $q\times R$ superposes both the continuum and the discrete descriptions of a given interaction. For the sake of simplicity, the discrete description is also represented by a smooth and continuous curve. The region delimited by the two curves and the demarcating points is, by definition, not resolved with the present technology. The two demarcating points represent the  experimental resolution thresholds of the two descriptions. They are, by definition, dependent of the existing technology but there is, inside this region, a critical point defined by the absolute equality of the two descriptions, which is technology independent. The existence of this critical point sets a scale for the interaction intensity in terms of an effective time derivative of $q(\tau)$. The discrete field formalism, we remind, being conformally symmetric \cite{hep-th/0006237} is scale free. 
\vglue13cm
\hglue-1.0cm
\begin{minipage}[]{7.0cm}
\parbox[t]{5.0cm}{
\begin{figure}
\vglue-13cm
\epsfxsize=250pt
\epsfbox{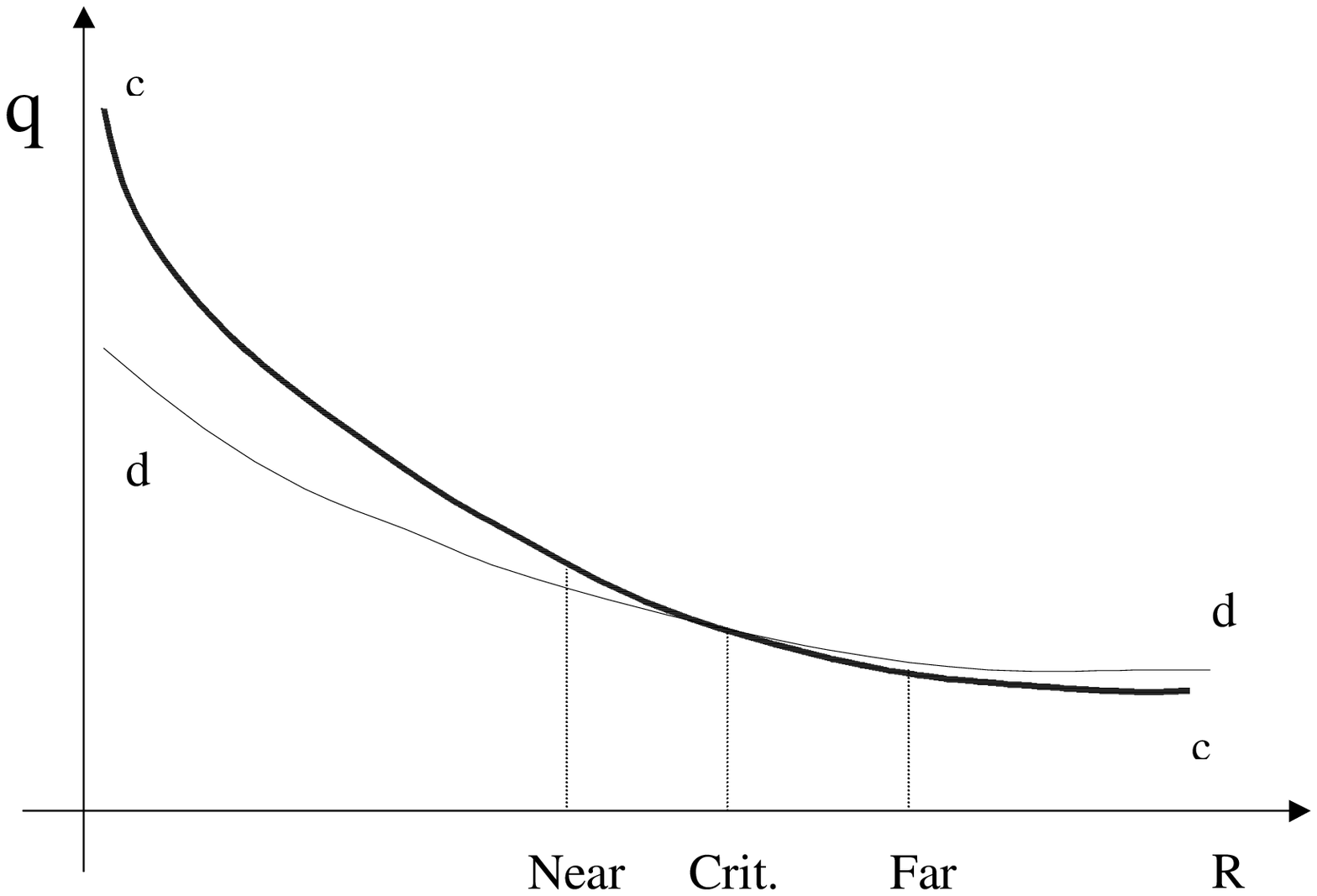}
\end{figure}}
\end{minipage}\hfill
\vglue-7cm
\hglue-1.0cm
\begin{minipage}[]{7.0cm}
\begin{figure}
\hglue8.0cm
\parbox[t]{6.5cm}{\vglue-6cm
\caption{Two descriptions for a same interaction: continuous (cc) and discrete (dd). For convenience the discrete one is approximated by a smoothly continuous curve. The near and the far demarcating points delimit the thresholds of existing technology for resolving the two curves. The critical point, defined by the two-curve crossing is technology independent and represents a fundamental scale for the interaction intensity in terms of an effective derivative of $q(\tau)$.}}
\end{figure}
\end{minipage}

\vglue-1cm

The two curves are just, respectively discrete and continuum, representations of a given generic interaction but their asymptotes, both near and far, carry universally valid  informations that make explicit and determinant the differences between the two descriptions. We are interested on these asymptotic regions where, in principle, discrepancies between them can be detected. An interaction in the near asymptotic limit where $\Delta\tau_{j}$ but not $\Delta q_{j}$ goes to zero diverges in the continuum description as $\frac{\Delta q_{j}}{\Delta\tau_{j}}$ goes to infinity whereas it remains finite in the discrete one. In the discrete description the interaction is always finite, no matter how strong. The resulting divergence indicates the inadequacy of the continuum formulation. This has been discussed in \cite{hep-th/9610028,hep-th/0006214,gr-qc/9801040} for both the gravity and the electromagnetic field: the inconsistencies of the continuum fields, made explicit through divergences and causality violations, disappear with the discreteness, with the existence of a non null lapse of time between two consecutive interaction points, or in other words,  with each interaction point being an isolated event. 

In the far asymptote, for an interaction with
\be
\Delta q_{j}=const\not=0,
\ee
${\dot q}$ in the continuum description goes to zero as $R$, and therefore $\Delta\tau_{j}$, goes to infinity whereas the discrete interaction tends to a finite and constant value; it just becomes sparser but does not go to zero. 
This demarcating point is for very large distances where $\Delta \tau_{j}$ may even become detectable so that the field asymptotic limit should reveal its discrete nature. Actually this possibility is spoiled, in the case of a matter-polarizing field like the electromagnetic one, by the shielding effect: the field is cancelled before $\Delta \tau_{j}$ grows to the point of detectability. This, of course, does not happen to gravity and so effects of this expected discreteness must be observable, but this discussion will be deferred to Section VI.

Careful observation at such small and large distances should reveal that the intensity of the actual interaction, respectively, grows and decreases at a smaller rate than the theoretical prediction from a continuous interaction. When observed, in a context of continuous interactions, these effects may require the use of regularization and renormalization techniques or may give origin to various misleading interpretations like the existence of new form of fundamental continuous interactions or of strange and yet to be observed form of matter, for example. In practice a detailed strictly discrete calculus is not always necessary and in some cases may  not even be feasible.
Calculus (integration and differentiation) in a discrete-interaction context becomes a useful approximation for a rigourous description of fundamental physical processes. 
 What effectively counts is the scale determined by $\Delta\tau_{j}$, the time interval between two consecutive interaction events, face the accuracy of the measuring apparatus. The question is if $\Delta\tau_{j}$ is large enough to be detectable, or how accurate is the measuring apparatus used to detect it. The density of interaction points on the worldline of a given point charge is proportional to the number of point charges with which it interacts. Let us consider the most favourable case of a system made of just two point charges. As the argument is supposedly valid for all fundamental interactions one can take the hydrogen atom in its ground state for consideration, treating the proton, for simplicity, as if it were also an elementary point particle. The order of scale of $\Delta\tau_{j}$ for an electron in the ground state of a hydrogen atom is can be taken en by the Bohr radius  divided by the speed of light
$$\Delta\tau_{j}\sim10^{-18}s$$
which corresponds to a number of ${\pi\over\alpha}\sim400$ interactions per period ($\alpha$ is the fine-structure constant) or $\sim10^{10}$ interactions/cm. So, the electron world  line is so densely packed with interaction events that one can, in an effectively good description for most of the cases, replace the graph of the Figure 2 by a continuously smooth curve. The validity of calculus in physics is then fully re-established in the interval between the two critical points as a consequence of the limitations of the measuring apparatus. The Wigner's questions\cite{Wigner} about the unexpected  effectiveness of mathematics in the physical description of the world is recalled.  The answer lies on the huge number of point sources in interaction, the large value of the speed of light and the small (in a manly scale) size of atomic and sub-atomic systems, which indirectly is a consequence of h, the Planck constant.

Even in these situations where $\Delta\tau_{j}$ may not be measurable, at least with the present technology, the discrete formalism is justified not for replacing the continuum one where it is best, which       is confirmed by high precision experiments \cite{Nakanish,Darmour}, but mostly for defining and explaining its limitations. There are, besides this very generic justification, many instances of one-interaction-event phenomena, like the Compton effect, particle decay, radiation emission from bound-state systems, etc, where discrete interactions are the natural and the more appropriate approach. These are, of course, all examples of quantum phenomena, but primarily because quantum here implies discreteness. 
\subsection{Discrete-continuum transition}
It would be interesting to have a framework where this change from continuum to discrete interaction and vice-versa could be formally realized in a simple and direct way.
One can deal with it considering the behaviour under a derivative operator of  ${\bar\theta}(\tau)$ which is the mathematical description of the interaction discreteness. Then one must require that, symbolically
\be
\l{dk}
\frac{\partial}{\partial\tau}{\bar{\theta}}(\tau-\tau_{i}):=\delta_{\tau\tau_{i}},
\ee
with $\delta_{\tau\tau_{i}}$ the Kronecker delta
\be
\delta_{\tau\tau_{i}}=\cases{1, &if $\tau=\tau_{i}$;\cr
	&\cr
	0, & if $\tau\ne\tau_{i}$,\cr}
\ee
with the meaning that at the points where the left-hand side of Eq. (\ref{dk}) is not null, which are the only relevant ones,  $\tau$ must be treated as a discrete variable and that the operator $\frac{\partial}{\partial\tau}$ must be seen as (or replaced by) just a sudden and finite increment and not as the limit of the quotient of two increments.

Then with such a convention one has from Eq. (\ref{dV}) that 
\be
\l{nabla}
\nabla_{\nu}q(\tau){\Big |}_{f}=-f_{\nu}\sum_{i}q_{\tau_{i}}\{{\bar\theta}(\tau_{i+1}-\tau)\delta_{\tau\tau_{i}}-\delta_{\tau_{i+1}\tau}{\bar\theta}(\tau-\tau_{i})\}:=-f_{\nu}{\dot q}(\tau),
\ee
which implies that ${\dot q}(\tau)$ is null when $z(\tau)$ is not a point of interaction on the charge world  line. For such an interaction point $\tau_{j}$ one has
\be
{\dot q}(\tau_{j})
=q_{\tau_{j}}{\bar\theta}(\tau_{j+1}-\tau_{j})-q_{\tau_{j-1}}{\bar\theta}(\tau_{j}-\tau_{j-1})=q_{\tau_{j}}-q_{\tau_{j-1}}
\ee
or, generically
\be 
\l{ac}
{\dot q}(\tau)=\cases{\Delta q_{i}=q_{\tau_{i}}-q_{\tau_{i-1}} & for $\tau=\tau_{i}$;\cr
		&\cr
		0 & for $\tau\ne\tau_{i}$,\cr}
\ee
and, from the middle term of Eq. (\ref{nabla}),
\be
\nabla_{\sigma}\nabla_{\nu}q(\tau){\Big |}_{f}=-2f_{\sigma}f_{\nu}\sum_{i}q_{\tau_{i}}\delta_{\tau\tau_{1}}\delta_{\tau_{i+1}\tau}=0.
\ee
 In Eq. (\ref{nabla}) $i$ labels the vertices and only these points on the world  line contribute to the sum. That is why one has to define  Eq. (\ref{dk}). In a limit where a summation over $i$ may be approximated by a line integration the Kronecker delta may be replaced by a Dirac delta function and then one may have Eq. (\ref{dasdd}) as a good operational approximation to Eq. (\ref{ac}).\\ 
Then
\be
\l{dphij}
\phi_{f}(x)=q(\tau){\Big |}_{f}=\cases{q_{\tau_{j}}{\Big |}_{f} &if $\tau_{j}<\tau_{ret}<\tau_{j+1}$\cr
		&\cr
		&\cr
		 \frac{q_{\tau_{j}}+q_{\tau_{j-1}}}{2}{\Big |}_{f}& if $\tau_{ret}=\tau_{j}$\cr}
\ee
\be
\l{dAj}
\nabla_{\mu}\phi_{f}(x)=-f_{\mu}\Delta q(\tau){\Big |}_{f}=\cases{-f_{\mu}(q_{\tau_{j}}-q_{\tau_{j-1}}){\Big |}_{f} &if $\tau_{ret}=\tau_{j}$\cr
		&\cr
		 0& if $\tau_{ret}\not=\tau_{j}$\cr}
\ee 
\be
\l{ddA}
\nabla_{\sigma}\nabla_{\nu}\phi_{f}(x,\tau)=0.
\ee
The meaning of Eqs. (\ref{dAj}) and (\ref{ddA}) is that the emission/absorption of a discrete field is an isolated event on its discrete-source worldline. Furthermore, the Eq. (\ref{ddA}) implies that only first-order field equations make sense for discrete fields. Higher order field equations, like the wave equation for example, have a meaning as an approximated description for a worldline densely packed with interaction-events (approximated interaction continuity) where the usual definition of derivatives may be used  and Eq. (\ref{ddA}) may be replaced by Eq. (\ref{dfe}). This points to an overall picture where the so called matter fields are fermions which are described by first-order equations whilst the fundamental (vectorial and tensorial) bosons are gauge fields whose meaning and origin, as discrete fields, we  discuss now. 
\subsection{The meaning of a gauge field}
It is worthwhile, after seing Eq. (\ref{ddA}), to recall how the discrete  field concept enlightens \cite{hep-th/9911233} the meaning and origin of gauge fields. What is behind a gauge symmetry?
 A discrete field is characterized by its (free) propagation without a change between two consecutive interaction events. A ``derivative" of a discrete field $\Psi_{f}$, which is a derivative along the field straight line support, is, therefore, identically null in this time interval. It is then, by definition, as a local function at a point $x$  
\be
\l{lim}
\nabla_{\mu}\Psi_{f}(x):=lim_{\epsilon^{\mu}\rightarrow0}(\Psi_{f}(x^{\mu}+\epsilon^{\mu})-\Psi_{f}(x^{\mu}-\epsilon^{\mu}))\equiv0,
\ee
an identically null difference if $x$ is not an interaction point. $\epsilon^{\mu}$ are parameters on the $f$-straight line support of $\Psi_{f}$.  See the Figure 4. This characterizes the field freedom, the absence of interaction in this time interval between two consecutive interaction events. The nullity on the RHS of the Eq. (\ref{lim}) is not, of course, valid at an interaction event. In order to have a general expression of everywhere validity a ``covariant derivative" must replace this derivative
\be
\l{ntoD}
\nabla_{\mu}\Psi_{f}(x)\Longrightarrow D_{\mu}\Psi_{f}(x):=\nabla_{\mu}\Psi_{f}(x)-A^{f'}_{\mu}(x)\Psi_{f}(x)\equiv0.
\ee
Eq. (\ref{ntoD}), that can easily be generalized for non-abelian fields, is a definition of the field $A^{f'}_{\mu}$ as the change in $\Psi_{f}$ attending the field conservation laws (energy-momentum, charge, etc.).
The product  $A^{f'}_{\mu}\Psi_{f}$ assures that this term is not null only at the interaction point, at the intersection of the straight lines $f$ and $f'$. See the Figure 4. 
\vglue-2cm
\hglue-2.5cm
\begin{minipage}[]{7.0cm}
\parbox[t]{5.0cm}{
\begin{figure}
\vglue1.5cm
\epsfxsize=250pt
\epsfbox{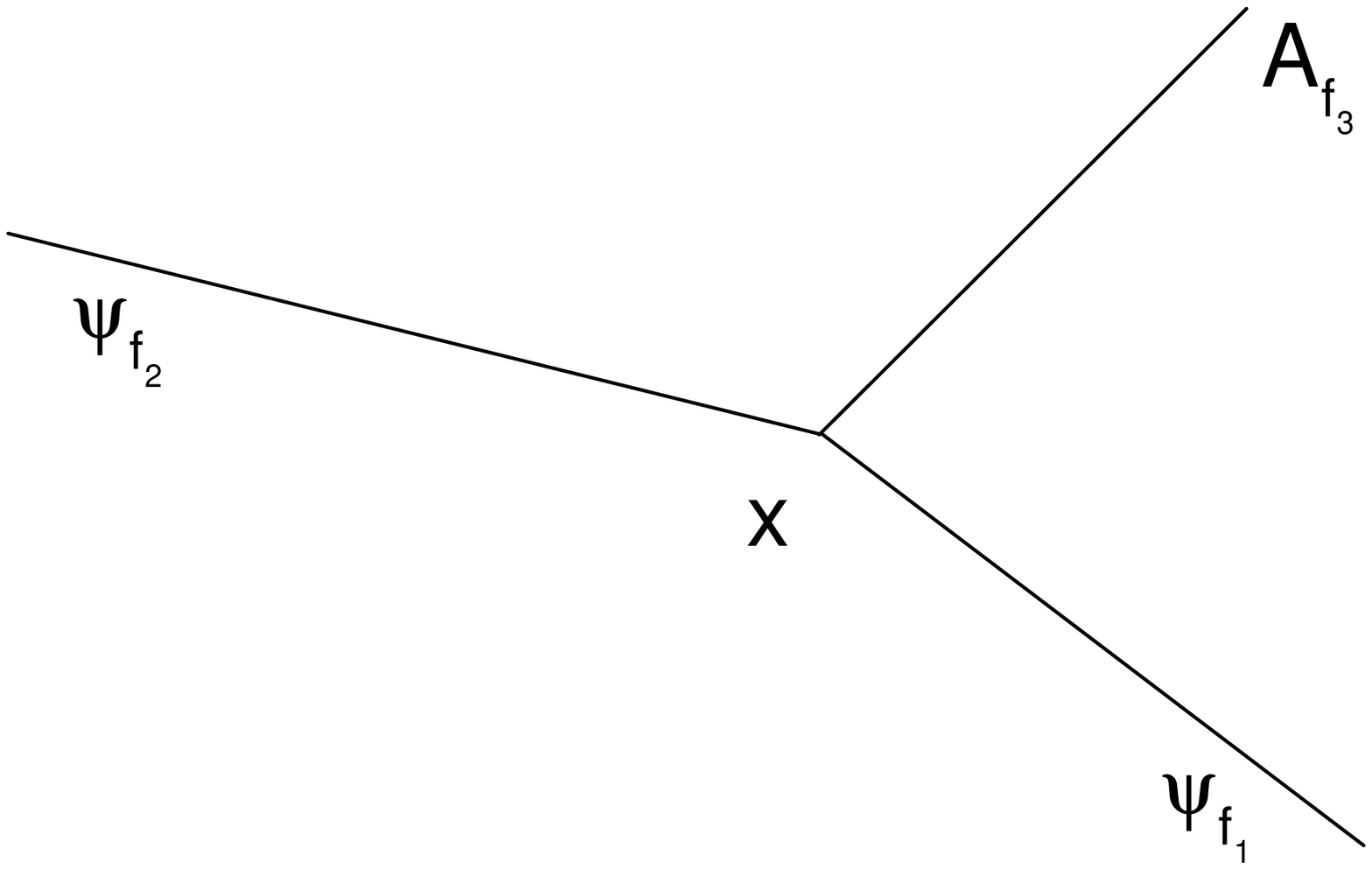}
\end{figure}}
\end{minipage}\hfill
\vglue1.5cm
\hglue-.5cm
\begin{minipage}[]{7.0cm}
\begin{figure}
\hglue7.5cm
\parbox[t]{7.0cm}{\vglue-7cm
\caption{The meaning of discrete gauge field and of covariant derivative. The covariant derivative of a discrete field is identically null as the field propagates without changing itself except at the interaction events where its associated discrete gauge field is its very change, as required by the field conservation laws. $f_{1},\;f_{2}$ and $f_{3}$ are three non-collinear four vectors.}}
\end{figure}
\end{minipage}

\vglue-3.5cm

The field $A^{f'}_{\mu}$ is ``gaugely"associated to $\Psi_{f}$: Either its creation is the consequence or its absorption is the cause of a change in $\Psi_{f}$; more than that, actually it is the very change. This very intuitive picture is not possible without the discrete field concept. Integration on the $f$'s replace the discrete fields by their continuum average fields retrieving the standard continuum field formalism. The effective continuum field $\Psi(x)$, in contradistinction to  $\Psi_{f}(x)$, changes its amplitude in the course of its propagation, and therefore, the right-hand side of the corresponding equations to (\ref{ddA}) and (\ref{lim}) are not identically null, as required in the continuum field formalism.
A discrete field, according to Eq. (\ref{dphij}), is just like an instantaneous picture of its source at its retarded time; a travelling picture.
If $z(\tau_{ret})$ is not a point of change in the source's state, the discrete field is not endowed with a physical  meaning as its energy tensor is null. A physical discrete field always corresponds to a sudden change in its source's state at its retarded time. If there is no change the field is not real, in the sense of having zero energy, zero momentum, zero charge, etc. Having no physical attribute it corresponds to the pure ``gauge field" of the continuum formalism. Moreover, Eqs. (\ref{lim}) and (\ref{ntoD}) imply that a gauge field cannot be a scalar field except in a strictly unidimensional interaction, with collinear four-vectors $f_{1}$, $f_{2}$ and $f_{3}$ of Figure 4. It excludes the existence of fundamental scalar fields.

\section{Scalar field and general relativity}
In this Section we discuss the unavoidable connections between the discrete scalar field and general relativity with its inherent notion of continuous interactions. 
That gravity be either totally \cite{inMTW} or partially \cite{Fierz,Darmour} described by a scalar (continuum) field is an old idea\cite{FierzePauli,Jordan,Brans-Dicke}. The exterior (vacuum) solutions of general relativity can be generated \cite{gr-qc/9801040} from a discrete scalar field in a metric theory but we will see now that this is not possible for solutions with massive sources as they do require a true metric field tensor. 

A positive change $\Delta q$ on the charge $q$ of a scalar source means that a scalar field $\phi_{f}(x,\tau)$ has been, say, absorbed whereas a negative one means then an emission. Therefore, a discrete scalar field carries itself a charge $\Delta q$ and can, consequently, interact with other charge carriers and be a source or a sink for other discrete scalar fields. It carries a bit of its very source, a scalar charge; it is an abelian charged field. 
 On the other hand a new look at equations (\ref{psf}) and (\ref{ac}) reveals that $(\Delta q_{j})^2$ describes the energy-momentum content of the field. So, energy-momentum is the source of a discrete scalar field, whose charge is a scalar function of energy.  Energy, of course, is a component of a four-vector and not a Lorentz scalar. Its four-vector character comes from the $f^{\mu}$ factor in Eq. (\ref{psf}). So, $\phi_{f}(x,\tau)$ is, in some way, associated to the gravitational field since they have a common source.
But, according to general relativity, gravity is a metric field. It is a measure of the spacetime curvature, of how it deviates from a flat spacetime; smoothness and continuity are inherent to general relativity. On the other hand, $\phi_{f}(x)$ is a discrete (pointlike) field; smooth and continuous is its spacetime average $\phi(x)$. Therefore, regarding $\phi_{f}(x)$ as a point-deformation in a Minkowski spacetime, propagating on a null direction $f$, the locally deformed geometry can be described by the following metric tensor
\be
\l{gmn}
g_{\mu\nu}^{f}(x)=g_{\mu\nu}{\Big |}_{f}=\eta_{\mu\nu}-\chi 
f_{\mu}f_{\nu}\phi_{f}(x,\tau).
\ee
This picture is validated if $g_{\mu\nu}^{f}(x)$ shows to be a solution of the Einstein field equations
\be
\l{Eeq}
R_{\mu\nu}-{1\over2}g_{\mu\nu}R=\chi T_{\mu\nu}
\ee
and if its spacetime average, in the sense of Eq. (\ref{s}) reproduces the corresponding Einstein solution $g_{\mu\nu}(x)$.   
With Eqs. (\ref{dAj}) and (\ref{gmn}) the harmonic coordinate condition $\partial_{\mu}g_{f}^{\mu\nu}=0$ is automatically satisfied and the Ricci tensor is expressed as 
\be
\l{Rmn}
R^{f}_{\mu\nu}:=R_{\mu\nu}{\Big |}_{f}=\frac{1}{2}f_{\mu}f_{\nu}\eta^{\alpha\beta}\nabla_{\alpha}\nabla_{\beta}\;\phi_{f}(x),
\ee
which is meaningful, we remind, only in the sense of an approximation of dense set of interaction events. 
Therefore, the Equations (\ref{Eeq}) are reduced to 
\be
\l{rEeq}
f_{\mu}f_{\nu}\eta^{\alpha\beta}\nabla_{\alpha}\nabla_{\beta}\;\phi_{f}(x,\tau)=\chi T_{\mu\nu}{\Big |}_{f}.
\ee 
It is remarkable that the highly non-linear Einstein Equations are reduced to a linear wave-equation. This is a consequence of discrete fields and of $f^2=0$. 
But Eq. (\ref{rEeq}) must be identified with Eq. (\ref{dfe}) for the discrete scalar field with
\be
T_{f}^{\mu\nu}=f^{\mu}f^{\nu}\rho_{f}(x),
\ee
as, from Eq. (\ref{rEeq}), $f_{\mu}T_{f}^{\mu\nu}(x)=0$ and $\eta_{\mu\nu}T_{f}^{\mu\nu}(x)=0$. There is no problem if $T_{f}^{\mu\nu}$ describes a massless source. For example, it can be shown that the Vaydia metric can be retrieved from a superposition of the discrete fields of a spherical distribution of massless dust \cite{unpublished}. For a massive source $T_{f}^{\mu\nu}$ could be, as discussed in \cite{hep-th/0006237}, a projection of 
\be
\l{rom}
T^{\mu\nu}_{m}=v^{\mu}v^{\nu}\rho_{m}
\ee
on the local lightcone, where $v$ is a four-velocity and $\rho_{m}$ is a pointlike-mass (a dust grain) density, but then there would be a problem of consistency. The point is that $\phi_{f}$ carries a little bit of its very source density $\rho$ which cannot be $\rho_{m}$, as $\phi_{f}$, being a gravity component, is a massless field. In other words, the source of gravity is not the mass but its energy-momentum; a graviton carries part of the energy-momentum but not of the mass of its source. This lets explicit a known symmetry of nature: the four fundamental interactions are described by gauge fields having vector currents for sources ($j=qv,$ as they are pointlike sources), of their respective charges $q$, including gravity since the energy tensor is just a current of its charge, the four-vector momentum. So, this symmetry is not broken with gravity being a second-rank tensor field and it has further consequences (see Eq. (\ref{emc})) that do not depend on the field tensor nature. 
\subsection{Flat or curve manifold?}
We are not proposing, it is worth repeating, the replacement of general relativity in its domain of validity by a discrete  field theory of gravity, and similar statements should be assumed  for other field theories. The geometrical description of gravity as the curvature of a pseudo-Riemannian spacetime has its validity, in the range of the ratio parameter (\ref{ratio}) limited by the two critical points, always assured as an absolutely good approximation due to the high density of interaction points in any real measurement. The point is that in this domain, i.e. for $\frac{\Delta q_{j}}{\Delta\tau_{j}}$ between the two demarcating points, there is, by definition, no experimentally detectable difference between a continuum and a discrete interaction descriptions. Considering the small strength of its coupling the gravitational interaction is irrelevant for physical systems involving relatively few fundamental elements. Even a gravitational Aharanov-Bohm-like experiment \cite{Sakurainaoda!} would require the gravitational field of a macroscopically large object, like the Earth. The sufficient condition for a high density of interaction points is assured and justifies continuous descriptions of gravity, of which general relativity seems to be the best proposal \cite{Darmour}. Moreover  the undetectability of discrete gravity in this region is tantamount to the unobservability of the Minkowski spacetime. At this level  of approximation the Minkowski spacetime becomes the local tangent space of a curved space-time and $f$ a generator of the local hypercone in the tangent space. This would lead to full general relativity in accordance to a general uniqueness result \cite{MTW,Visser} that any metric theory with field equations linear in second derivatives of the metric, without higher-order derivatives in the field equations, satisfying the Newtonian limit for weak fields and without any prior geometry must be exactly Einstein gravity itself. 

The Eq. (\ref{gmn}) reminds an old derivation \cite{Feynman} of the field equations of general relativity by consistent re-iteration of 
\be
\l{reit}
g_{\mu\nu}(x)=\eta_{\mu\nu}+\chi h(x)_{\mu\nu},
\ee
as solution from a gauge invariant wave equation for the field $g_{\mu\nu}(x)$ in a Minkowski spacetime. The non-linearity of the Einstein's equations comes from contribution to $g_{\mu\nu}(x)$ from all terms of higher orders in $h_{\mu\nu}$. For
$h_{\mu\nu}=f_{\mu}f_{\nu}\phi_{f}(x,\tau),$ there is no higher order contribution essentially because $f^2=0$. A discrete field has no self-interaction, a consequence of its definition (\ref{df}) and that is explicitly exhibited in its Green's function (\ref{pr9}). Discrete fields are solutions from linear  equations. Whereas this is true for $g^{f}_{\mu\nu}$ of Eq. (\ref{gmn}) it is not for its $f$-averaged $g_{\mu\nu}$ of Eq. (\ref{reit}). The non-linearity of general relativity appears here then as a consequence of the averaging process of Eq. (\ref{s}) that effectively smears the discrete field over the lightcone, erasing all the information contained in $f$. 
The approach of starting from Eq. (\ref{reit}) for deriving general relativity from flat spacetime receives now the distinctive aspect that the effective Riemannian spacetime comes not from a consistency requirement but as an approximation validated by a recognized limitation on our experimental capacity, which can always be improved, be placed on more stringent limits, but never be totally eliminated.

\section{Conclusions}
 
The thesis that all fundamental interactions are discrete is being developed. If this is the case there is no really compelling reason for excluding gravity from such a unifying idea.
The knowledge of a supposedly true discrete character of all fundamental interactions is a permanent reminder of the limits of a continuum approximative description. There are many indications of possible theoretical and observational evidences of direct consequences of interaction discreteness, particularly in gravity, of which we can just make some brief comments, although they are full-attention deserving. 
Discrete interactions imply on the ratio parameter (\ref{ratio}) whose range is divided in three segments by its two (the far and the near) critical values. The interior segment, between these two values, defines the domain of validity of the continuous-interaction approximation where the polygonal worldline of the sources are so densely packed of interaction points that they can be effectively replaced by smoothly continuous curves where the concept of acceleration and of spacetime curvature at a point on the worldline make sense. 

On the other hand, for values of the ratio parameter (\ref{ratio}) in the exterior segments, i.e. below or above the two critical points the discrepancies between a discrete and a continuous interaction cannot be overlooked. These are the places  where great riddles are associated to gravity. They are not easily fitted with a continuous interaction without evoking new forms of interaction or of (dark) matter and energy. This casts doubts on the results about asymptotical fields and their singularities of any continuum-field theory. 
\subsection{Large distance effects}
Gravity, in contradistinction to all other fundamental interactions has problems in the large distance limits where the field is extremely weak. The continuum asymptotically null fields are replaced by discrete interactions that become more and more sparse with the distance. This may be detectable for the gravitational field as it does not have shielding effects although it requires huge masses for detecting very weak gravitational fields and huge distances for producing a detectable $\Delta\tau_{j}$; both conditions found at and above galactic scales.  Therefore, a right place for  checking for signs and applications  of  discreteness is on dynamics of galaxies and on cosmology. Flat rotation curve of galaxies, inflation, accelerated expansion may be just manifestation of a discrete gravity. Asymptotic flat rotation curves appear very naturally \cite{gr-qc/0103218} in a discrete-field context!  It is therefore a real possibility that the critical point for gravity has already been detected in the flat rotation curves of galaxies \cite{rotation curves of galaxies}. The flatness feature of a rotation curve of a galaxy, as remarked by Milgrom \cite{Milgrom}, is determined not by its central mass $M$ alone nor just by the distance $R$  but by the acceleration which is equivalent to the ratio-parameter (\ref{ratio}) as $\Delta q_{j}$ corresponds to a change of speed. Therefore the existence of the critical point in the continuum/discrete physical description justify the introduction of a new fundamental scale for effective accelerations, and may put Milgrom's MOND on a sound physical basis. 
 The actually prevailing wisdom that a flat rotation curves is the (ad hoc) indication of some strange, ubiquitous but still to be detected cold dark matter is not free of problems and is far from being unanimous\cite{Milgrom,cdm,Evans,Mannheim,Nucamendi}. 

A gravitational repulsion, instead of attraction may be just a residual correction of the excess committed in the continuum approximation. In a scale of distance where the effective attractive interaction is so weak that sub-dominant contributions  to discrete gravity dominate \cite{gr-qc/0111152}.

Another possible evidence of discrepancy that must be considered is the apparent anomalous, weak, long-range acceleration observed in the Pioneer 10/11, Galileu, and Ulysses data \cite{Pioneer}. Due to their spin-stabilization and to the great distance (30 t0 67 AU) from the Sun the spacecraft are excellent for dynamical astronomy studies as they permit precise acceleration estimation to the level of $10^{-10}cm/s^{2}.$ The detected anomalous acceleration comes from neglected second largest contribution from those mentioned $n$-combinatorials, although it is still not clear if this anomaly is not a too large contributions from discrete interactions. 
\subsection{Small distance effects}

In the asymptotical short-distance limit, below the critical point, the fact that the discrete field is weaker than the continuum one suggests a natural explanation to inflation in cosmological theories.

Other possible indication of discrete physics are the experimentally observed (or just realized) manifestations of low (one- or two-) dimension physics. It is an enormous difficulty for the continuum field formalism to explain a mechanism for restricting the interaction field to a one- or two-dimensional space instead of spreading over the whole three-dimensions.   Quantum Chromodynamics faces similar problems for explaining quark-confinement. 
Solid state physics is, in a discrete interaction context, a totally unexplored subject, rich of such possible manifestation of discrete physics. Possible examples are the quantum Hall Effect and the high $T_{c}$ superconductivity where indications of two-dimensional physics have been observed. Crystal or molecular structural arrangements constraining the movement of the relevant charges to layers (subdimensional regions) restrict, as a consequence, the discrete, but not the continuum, interaction fields to these sub-regions. Whereas this is easily explained with discrete fields it does not work for the continuum one as it would require ultrarelativistic velocities for an approximate (but in the wrong space direction) dimensional reduction of the field 3-dimensional support. This same mechanism explains, in terms of discrete fields, light polarization and the action of a polaroid.  These constraints may be dependent on the conditions of low temperatures with its consequent suppression of some thermal excitations.

Since there is no differential equations and neither integrations with discrete interactions, the evolution of any system is done through successive, sudden and discrete finite differences that are just super imposed. Between two consecutive interaction points every pointlike component just moves freely on straight lines. It is then remarkable that the whole continuum-field physics is reproduced as an effective approximation after a large number of discrete interactions \cite{gr-qc/0111152}. Any exact physical statement must be expressed as finite power series of combinatorials of the accumulated number of interaction events.  So the world is surprisingly simpler and our standard vision of it is richer of such idealized, unreachable concepts than we had previously conceded. A whole paraphernalia of mathematical tools, so useful in physics - differential equations, integrations, differential geometry, topology, just for citing few -  and also so many familiar and daily used mathematical functions like sine, exponential, harmonic and coulombian potentials, and circles, ellipse, etc, do not belong to the realm of the physical world; they are just unreachable limiting concepts as much as an ideal gas and a reversible process. This is reflected in the
 Tsallis's \cite{Tsallis,Tsallis's homepage} generalized one-parameter statistics based on a power-law distribution of probabilities. It is reduced to Boltzmann statistics when its parameter is equal to one. This parameter is then a measure of how close the system is from its idealized asymptotic state, that rigorously, is reachable only after an infinite number of interactions. It may be an appropriate statistics for a world made of sets of discretely interacting pointlike objects.
\subsection{Concluding remarks}
The idea of an essential continuity of any physical interaction allows unlimited speculations that will always go beyond any level of possible experimental verifications which brings then the risk of not being able of distinguishing  the reign of possibly experimentally-grounded scientific research  from plain philosophical speculation or even just fiction. Regardless the possibility that some of its consequences have already been experimentally detected, a discrete gravitational interaction, even in other range where it is not experimentally detectable, still for a long time to come, may just make sense of existing theories for delimiting their domain of validity  as it has historically happened with all new discreteness introduced in the past, like the ideas of molecules, atomic transitions, and quarks, for example.

\end{document}